\documentclass[prd,reprint,letterpaper,superscriptaddress,amsfonts,amsmath,amssymb,nofootinbib]{revtex4-2}

\usepackage{braket}

\usepackage{graphicx}
\usepackage{color}
\usepackage[normalem]{ulem}
\usepackage{MnSymbol}
\usepackage{enumerate}
\usepackage{bbold}
\usepackage{subfigure}
\usepackage{subfiles}
\usepackage{units}
\usepackage[usenames,dvipsnames]{xcolor}
\usepackage[colorlinks=true,linkcolor=MidnightBlue,citecolor=blue,filecolor=green,urlcolor=Violet]{hyperref}
\usepackage[dvipsnames]{xcolor}
\usepackage{paralist}
\usepackage{psfrag}
\usepackage{setspace,tabularx,fancyhdr}
\usepackage{mathtools}
\usepackage[top=2cm,bottom=2cm,left=2cm,right=2cm]{geometry}
\usepackage{enumitem}
\setlist[itemize]{leftmargin=*}
\usepackage{wrapfig}
\usepackage{type1cm}
\usepackage{yfonts}

\def\beq{\begin{equation}}
\def\eeq{\end{equation}}
\def\bsp{\begin{split}}
\def\esp{\end{split}}
\def\bea{\begin{eqnarray}}
\def\eea{\end{eqnarray}}

\def\Oprr{\hat{\mathbf{r}}}
\def\Opromeg{\hat{\mathbf{\omega}}}

\def\cosp{\cos{\phi}}
\def\cospp{\cos^2\phi}

\def\AvR{\Bar{R}}
\def\AvOme{\bar{\mathbf{\omega}}}

\def\GothA{\textgoth{A}}

\newcommand{\neweqline}{\nonumber\\}
\def\Avx{\bar{x}^j}

\def\K{\mathbf{K}}

\definecolor{myyellow}{rgb}{0.94, 0.86, 0.51}

\definecolor{mygreen}{rgb}{0.2, 0.8, 0.2}

\definecolor{mypink}{rgb}{0.99, 0, 0.99}

\definecolor{mypurple}{rgb}{0.75, 0, 0.75}

\definecolor{cadmiumorange}{rgb}{0.93, 0.53, 0.18}

\newcommand{\IGNORE}[1]{}

\newcommand{\kp}{\kappa}

\newcommand{\lp}{\ell_p}

\def\oR{\widehat{R}}
\def\bR{\hat{\mathbf{R}}}
\def\bxtwo{\hat{\mathbf{x}}_2}
\def\xtwo{\hat{x}_2}
\def\bxone{\hat{\mathbf{x}}_1}
\def\bytwo{\hat{\mathbf{y}}_2}
\def\byone{\hat{\mathbf{y}}_1}
\graphicspath{{Images/}}
\def\unitx{\mathbf{e}_x}
\def\unity{\mathbf{e}_y}
\def\ox{\hat{x}}
\def\oy{\hat{y}}
\def\oz{\hat{z}}
\def\oR{\hat{R}}

\begin{document}

\title{\texorpdfstring{$\kappa$-General-Relativity}{kappa-General-Relativity} II: 
An Astrophysical Observable from a 2-Body system}

\author{Daniel Rozental}\email{daniel.rozental@biu.ac.il}
\affiliation{Department of Physics, Bar Ilan University, Ramat Gan 5290002, Israel}

\author{Ofek Birnholtz}
\affiliation{Department of Physics, Bar Ilan University, Ramat Gan 5290002, Israel}

\begin{abstract}
We examine gravitational waves (GWs) from Binary Black Holes (BBH) as possible suitable systems for investigating the physical validity of theories predicting the Relative Locality (RL) effect, an effect arising in the $\kappa$-Minkowski non-commutative spacetime, a central property in theories of Quantum Gravity (QG) Phenomenology.
Hence, we are taking a step towards realizing the purpose of the phenomenological effort of having observational evidence to put constraints on QG approaches.
In particular, we show that the RL effect induces an uncertainty in the observed rotational frequency $\Opromeg$ during the inspiral phase. This uncertainty becomes stronger with increasing observational distance. It also increases with decreasing orbital radius, and it \textit{statistically accumulates} as an increasing variance of $\Opromeg$ over successive cycles. 
In terms of the post-Newtonian deviations, the uncertainty contributes at 1.25th order, an order that has not yet been directly constrained in GW analyses.
\end{abstract} 
\maketitle

\section{Introduction}\label{ssection: Introduction}
Following Paper I \cite{Our_Paper}, the main assumption underlying this work is that the description of events in terms of classical variable spacetime points should be abandoned when both gravitational and quantum effects become significant—namely, in regimes where Quantum Gravity (QG) is expected to play a crucial role.
We also restrict ourselves to a description based on a phenomenological model that allows for meaningful predictions. A promising approach in this direction is the $\kappa$-Minkowski spacetime, introduced in \cite{Bicrossproduct_Structure,q_deformations, Deformed_Quantum_Relativistic_Phase_Spaces, Kappa_Deformed_Phase_Space}. This framework describes events via a Non-Commutative Spacetime (NCST), governed by the operator relation\footnote{We use Greek letters ($\mu,\nu$, etc.) for spacetime indices $0,1,2,3$, and Latin letters ($i,j$, etc.) for purely spatial indices. $G_N$ denotes the universal gravitational constant, $\hbar$ is the reduced Planck constant, and $c$ is the speed of light; we use Quantum Field Theory units, where $\hbar=c=1$.}:
\beq\label{Kappa minkowski}
[\ox^0,\ox^{i}]= \frac{i}{\kappa}\ox^{i}, \quad \quad [\ox^{i},\ox^{j}] = 0,
\eeq
where $\kappa$ represents a characteristic energy scale, typically identified with the Planck energy, $\kappa = E_p = \hbar c/l_p \sim 10^{19} \, \text{GeV}$ and is proportional to the inverse of the Planck length, $l_p = \sqrt{\frac{\hbar G_ N}{c^3}} \sim 10^{-35} \, \text{m}$.

The connection between \eqref{Kappa minkowski} and QG phenomenology arises from the study of a family of theories known as Deformed Special Relativity (DSR). In DSR, one postulates that at the crossover between classical and QG energy scales, an imprint of quantum gravitational effects manifests as an invariant scale (in addition to the speed of light), such as a minimal length or maximal energy density 
\cite{Relativity_with_length1, quanta_mass}, while still preserving the principles of relativity (i.e., no preferred frame). By now, it is well established that for a DSR framework to be relativistically consistent, the symmetries of Special Relativity (SR) must be deformed accordingly, particularly in the interplay between the Lorentz sector and the translations \cite{Relativity_with_length2, quanta_mass}. 

It was soon recognized \cite{Different_Realization} that each DSR model’s symmetry structure can be associated with a distinct basis of the $\kappa$-Poincar\'e group, the symmetry group that preserves the invariance of the NCST \eqref{Kappa minkowski} \cite{Bicrossproduct_Structure, q_deformations, Deformed_Quantum_Relativistic_Phase_Spaces, Kappa_Deformed_Phase_Space}. This observation led to the conclusion that $\kappa$-Minkowski spacetime serves as the natural spacetime framework for DSR theories \cite{kappa_to_DSR}.

Relative Locality (RL) is a proposed phenomenon that can be linked to DSR and is deeply related to the deformation of symmetries and the NCST \eqref{Kappa minkowski}.
It states that locality, when constructed by highly energetic probes, is not an absolute notion, and that a collision of particles should be treated as local only for a local observer, whereas for distant ones locality is not guaranteed \cite{Relative_Locality}.
Such an effect is shown, with the help of the underlying $\kappa$-Poincar\'e deformed symmetry, to be relativistically invariant \cite{Relative_And_kappa_1, Relative_And_kappa_2, Boost_2,Our_Paper}, and can also be derived from the uncertainty that comes from the NCST itself \eqref{Kappa minkowski} 
\beq\label{Uncertainty Relation} 
\sigma_{\ox^{0}}\sigma_{\ox^{j}} \geq \frac{1}{2\kappa}\Avx,
\eeq
where $\sigma_{\hat{g}}$ represents the typical uncertainty in a measurement of the operator $\hat{g}$ relative to its averaged value $\bar{g}$. Such an uncertainty \eqref{Uncertainty Relation} implies a \textit{fuzziness} in the coordination of events: while a local observer can sharply coordinate events ($x^j=0$), a distant observer perceives the same events as fuzzy, representing a \textit{measurement} limitation for non-local observers, resulting in the loss of absolute locality typical of RL (see \cite{NonCommutative_Spacetime_Interpertation_3, NonCommutative_Spacetime_Interpretation1}).

Having reviewed the foundations and theoretical implications of the $\kappa$-Minkowski NCST, the next natural step is to explore its phenomenology—namely, possible ways to verify or falsify the assumptions regarding QG imprints predicted by DSR theories. Proposed tests include time delays between low- and high-energy particles, as well as anomalies in particle production, either arising from a deformed energy-momentum relation, a modified momentum addition law, or the underlying deformed symmetries (see, e.g., \cite{RL_Test_1, In_Vacua_Dis, RL_Test_2}). Other studies focused on how the NCST affects field theories constructed on it, such as scalar and quantum electrodynamics fields, or, as was the subject of our previous paper \cite{Our_Paper}, how general relativity gets deformed when \eqref{Kappa minkowski} becomes the local spacetime. We refer to \cite{Multy_Mass} for an extensive review on the observational effort in DSR. 

In this work, we focus on examining a new physical observable that originates from the uncertainty relation \eqref{Uncertainty Relation} and the RL principle, both of which are direct consequences of the NCST \eqref{Kappa minkowski}. Importantly, this observable depends only on the four-dimensional operator-valued spacetime realization of $\kappa$-Minkowski\footnote{Note that other types of NCST representations, such as Moyal-type constant NC or commutation relations of the form $[\hat{x}^\mu,\hat{x}^\nu] \sim C^{\mu\nu}_{\rho\sigma\ldots} \hat{x}^\rho \hat{x}^\sigma \cdots$, are not considered here. It would be of interest to investigate whether similar effects arise in those frameworks (note that the specific dependence in $\hat{\vec{x}}$ of the NCST relations is crucial in our construction).} \eqref{Kappa minkowski}, leading to the uncertainty relation \eqref{Uncertainty Relation}. This implies that the observables constructed directly from \eqref{Uncertainty Relation} do not depend on the specific models of phase space construction (i.e., the deformed symmetry) within the DSR framework that preserve the same spacetime algebra \eqref{Kappa minkowski} (see \cite{Different_Realization} for different deformed symmetries with the same $\kappa$-Minkowski structure), unlike, for instance, the deformed energy-momentum relation.

To determine how the uncertainty relation \eqref{Uncertainty Relation} can manifest in a physical system, it is useful to examine the study in \cite{NonCommutative_Spacetime_Interpretation1}. There, it was demonstrated (see Sec.~\ref{ssection: Observation Preliminaries} for a detailed discussion) that for an observer to recognize the fuzziness in the coordination of an event, they must compare their recorded coordinates with those of another observer at a different spatial separation from the event.  
Here, we adopt \cite{NonCommutative_Spacetime_Interpretation1}'s perspective but consider the comparison of event coordinates recorded at different \textit{times} by a single observer.
This leads us to propose, to the best of our knowledge, the first potentially observable phenomenon that directly stems from RL and \eqref{Uncertainty Relation} and that might be within reach of foreseeable technological capabilities. 

The outline of the paper is as follows: In Sec.~\ref{ssection: Observation Preliminaries}, we outline the necessary conditions a system must satisfy to serve as a candidate for observing the effects of \eqref{Uncertainty Relation}.
In Sec.~\ref{ssection: outline of the system}, we introduce and set the binary black hole system as a possible candidate.
In Sec.~\ref{Ssection: Some Definitions and notations} we establish some relevant formulae and notations, and eventually present the corresponding calculations of the observable quantities in Sec.~\ref{ssection: evaluation of uncertainties}. In Sec.~\ref{Observational sign}, we analyze the magnitude of the effect and assess its detectability using current and possible future observational technologies.
Finally, we summarize our findings and discuss future research directions in Sec.~\ref{ssection: summary and discussion}.

\section{Observation effects}
\subsection{Some Preliminaries}\label{ssection: Observation Preliminaries}
When studying the spacetime \eqref{Kappa minkowski} solely as a description of events (i.e., spacetime points) which exhibit the uncertainty relation \eqref{Uncertainty Relation}, one can learn about the physical behavior of $\kappa$-Minkowski, even without establishing a theory of dynamics on and of the spacetime.
The work done so far in such a direction includes a proof that it is consistent as a relativistic theory (see \cite{Our_Paper} for further details and citations).
However, we do not yet know of any obvious physical observation testing phenomenon of such uncertainty relation \eqref{Uncertainty Relation} directly, an observation with crucial insight on the NCST \eqref{Kappa minkowski}.

A significant step towards this goal was taken in \cite{NonCommutative_Spacetime_Interpretation1}, which, as discussed in Sec.~\ref{ssection: Introduction}, provided a framework for constructing observables in the NCST \eqref{Kappa minkowski} by treating it as a quantum spacetime, where time and space are operators in a Hilbert space. This approach allowed the authors to determine how the spatial and temporal uncertainty of a given event varies between observers at different distances. Consequently, it offered insights into how the RL effect manifests in a simplified observational scenario. However, to our knowledge, no concrete physical observable has yet been proposed to directly measure the RL effect.

In this section, we aim to extend the current theoretical framework of physical observables toward identifying a measurable physical phenomenon that arises from the uncertainty relation \eqref{Uncertainty Relation}. This also implies that we are neglecting possible \textit{generation}-induced deformations, as will be clarified shortly. 
\\\\
\begin{center}
\textbf{Conditions for Observing RL effects}
\end{center}
To detect the effect of the uncertainty relation \eqref{Uncertainty Relation} between the typical uncertainties of time and space, we need to consider a system that exhibits two key qualities, which, while framed differently, are similar to those outlined in \cite{NonCommutative_Spacetime_Interpretation1}:
\begin{itemize}
    \item First, the event must involve some coupling between space and time, ensuring that we can not alter one (e.g., shifting the origin) so that the uncertainty can be arbitrarily adjusted.
    This constraint can in fact be phrased as requiring that there is no time killing vector for the system, similarly to the known property that no deformation will appear in $\kappa$-GR when considering static spacetimes.
    \item Second, for the effect to be measurable, there must be intrinsic information about the system’s properties unaffected by the uncertainties.
    A suitable observational setup would allow us to probe the space-time uncertainty (as depicted in Fig.1 in \cite{NonCommutative_Spacetime_Interpretation1}). 
\end{itemize}

Thus, a static or stationary\footnote{In fact, "stationary" suffices here, i.e. the observed system possesses a time-killing vector field.} astrophysical target does not satisfy either of these two conditions. Although such a target might initially seem promising due to the large distance from the observer (on Earth), which results in uncertainty scaling linearly with the observer's distance operator $\bar{x}^j$, it fails to meet either of the requirements.

Instead, we propose a physical system that fulfills both conditions and provides an observable effect from the uncertainty relation \eqref{Uncertainty Relation}. Importantly, for this analysis, we assume that the \textit{generation} mechanism of the system remains classical, without corrections from $\kappa$-GR evaluated in \cite{Our_Paper}.
This can be treated as a first-order approximation in the size of the system, considered small compared to the distance to the observer, i.e. the generation is local. In such an  approximation, the $\kp$-Minkowski corrections are applied exclusively to the observational sector, while the system’s dynamics is dictated from GR. Alternatively, the observation effect can be considered in isolation. Note that the results presented here will still hold when incorporating corrections from $\kappa$-GR, provided the system exhibits certain symmetries, as discussed in \cite{Our_Paper, Metric_Perturbations_In_NC, NC_Geometry3}.
\\

\subsection{Outline of the system}\label{ssection: outline of the system}
We consider two black holes with masses $(M_1, M_2)$, respectively (the reason for focusing on black holes will become apparent later in Sec.~\ref{Observational sign}), orbiting each other in quasi-circular motion with a separation radius $\mathbf{R}$, with the phenomenological approximation that dynamics is dictated by \textit{classical GR}.
The usual expression gives the angular velocity of the binary system in the Newtonian/
Keplerian limit, to leading order:
\beq\label{local frequency} 
\omega = \sqrt{\frac{G_N M_T}{R^3}},
\eeq
where $M_T \equiv M_1 + M_2$ is the total mass, and $R=|\mathbf{R}|$.
This setup is commonly called a Binary Black Hole (BBH) system. The evolution of such a system can be divided into three main stages \cite{Basic_Physics, thorneLectures}.
In the first stage, the \textit{inspiral}, the black holes orbit each other in a stable manner, which for simplicity we take to be quasi-circular.
As energy is lost to gravitational radiation emission, the orbital energy decreases, causing the separation radius to shrink over time.
However, in the early stages of the inspiral, where $\partial_t(R) \ll \omega R$, the separation radius and the corresponding angular frequency can be treated as nearly constant over several orbital cycles (see \cite{GW_Binary_1, GW_Binary_2, Basic_Physics}, and Ch.4 in \cite{Maggiore:2007ulw}).
As time progresses, the separation radius continues to decrease due to the energy loss, reaching the $R_{\text{ISCO}}$ radius. At this point, the approximation of quasi-circular orbits breaks down, and the strong-field effects of GR become non-negligible.
This is the second, or  \textit{merger} stage.
Finally, during the \textit{ringdown} stage, the merged remnant of the two black holes settles down to a single final black hole, which is less massive than $M_T$ due to the loss of energy and angular momentum.

In this analysis, we focus on the \textit{inspiral} stage of the binary, where the frequency and separation radius evolve slowly enough to be considered constant over a small number of cycles.
We define the separation radius $\mathbf{R}$ and the angular frequency $\omega$ from \eqref{local frequency} as \textbf{local quantities}.
By \textit{local} quantities, we mean quantities \textit{measured in a reference frame local to the center of mass of the binary system}. Whereas \textit{local frame} refers, in this context, to a frame that is held fixed relative to outgoing infinity (e.g., to earth), and that is close enough to the center of mass of the BBH such that the right-hand side in \eqref{Uncertainty Relation} $l_p\bar{x}^j$ (with $\bar{x}^j$ taken as the distance from each mass to the local observer)
can be considered to vanish for any reasonable sensitivity\footnote{In fact, one could define the local quantities as the classical quantities where $l_p\rightarrow0$, or by a measurement in the early stage of the inspiral where $\mathbf{R}$ is large (see Sec.~\ref{ssection: protocol}), and the uncertainty is marginal.}.   

Now, consider an observation conducted from Earth. Incorporating the effect of the uncertainty relation \eqref{Uncertainty Relation}, we first posit that the measured quantities are now \textbf{operators}. They are defined as follows: $(\Oprr_1, \Oprr_2)$ represent the spatial locations of each black hole 
, $\bR$ denotes the measured separation radius vector $|\Oprr_1 - \Oprr_2|$ (boldfaced Latin letters stand here for vectors), $\hat{R}$ denotes the absolute value of $\bR$, and $\Opromeg$ represents the measured frequency. 
According to the space-time uncertainty \eqref{Uncertainty Relation}, each operator will exhibit some typical uncertainty from its average value.
For instance, for the operator $\Opromeg$, we have $\Opromeg=  \bar{\omega}  + \sigma_{\Opromeg}$, where we use $\sigma_{\hat{g}}$ to denote the variance of a scalar operator $\hat{g}$ as in \eqref{Uncertainty Relation}, and the (minimal) typical uncertainty should be dictated from the relation on the coordinates \eqref{Uncertainty Relation}. 
For the current discussion, we shall work here with three Euclidean dimensions so that the (operator) location of each black hole constructed by the observer takes the form
\bea\label{Euclidean coordinates}
\Oprr_1&=&\bxone+\byone=\ox_1\cdot\unitx+\hat{y}_1\cdot\unity+\oz_1\cdot\mathbf{e}_z,
\neweqline
\Oprr_2&=&\bxtwo+\bytwo=\ox_2\cdot\unitx+\hat{y}_2\cdot\unity+\oz_2\cdot\mathbf{e}_z,
\eea
where the $x$-axis is taken to be along the observer's line of sight and $(y,z)$ perpendicular to $x$ and to each other. Additionally, $(\unitx,\unity,\mathbf{e}_{z})$ are the basis vectors along the $(x,y,z)$ directions respectively. 

\begin{figure}[htbp]
    \centering
    \includegraphics[width=0.75\linewidth]{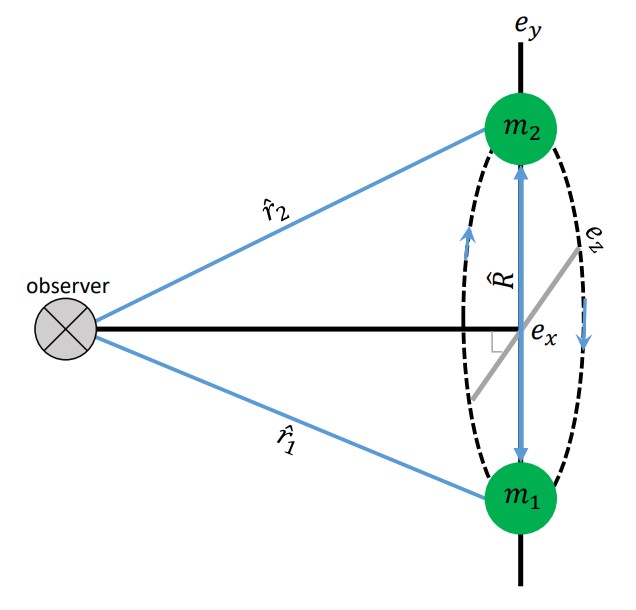} 
    \caption{Face on orientation: Case $(\parallel)$}
    \label{fig: face on}
\end{figure}

\begin{figure}[htbp]
    \centering
    \includegraphics[width=0.85\linewidth]{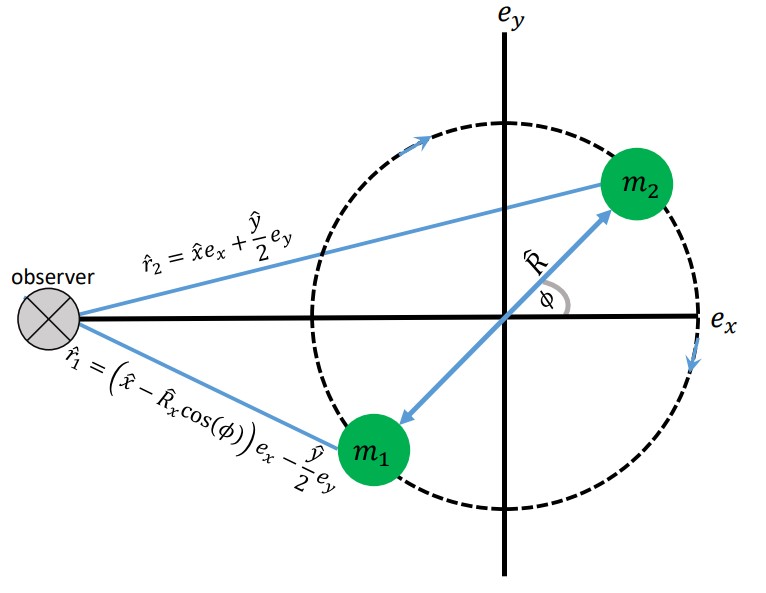} 
    \caption{Edge on orientation: Case $(\perp)$
    }
    \label{fig: edge on}
\end{figure}

Our aim is now to evaluate how the uncertainties coming from \eqref{Uncertainty Relation} will affect the observation of the gravitational waves coming from the BBH. For that, we notice that a BBH might rotate in different orientations relative to the observer; as a first step, we shall now consider two of them: 
\begin{itemize}
 \item \textbf{Face-On orientation $(\parallel)$:} the plane of rotation is along the $\hat{y}$-$\oz$ axes, both perpendicular to the line of sight axis $\ox$, while the axis of rotation is parallel to the line of sight, as illustrated in Fig.\ref{fig: face on}. The important aspect here is that the absolute value of the distance vectors is the same throughout the circular motion $|\Oprr_1| = |\Oprr_2|$ (note the assumption of equal masses, see the discussion around eq.\eqref{coordinate of the parallel case}). Another point here is that concerning the observer, the dynamics along $\hat{y}$ and the $\oz$ axes are the same, and we shall use this later. 
    \item \textbf{Edge-On orientation $(\perp)$:} the plane of rotation is along the $\ox$-$\hat{y}$ axes, so that $\ox$ is along the observer's line of sight and the axis of rotation is perpendicular to the line of sight. We denote the angle of the circular motion--between the $\ox$-axis and $\bR$ as $\phi\in[0,2\pi]$, as shown in Fig.\ref{fig: edge on}. The important aspect here is that the circular rotation varies the distance along the $\ox$-axis of each mass, and reaches maximal separation (absolute) between the masses when $\phi=0,\pi$. 
\end{itemize}

We will show in Sec.~\ref{ssection: evaluation of uncertainties} that for the ($\parallel$) scenario, the uncertainty in the measured magnitude of the separation radius $\oR^{\parallel}$ is marginal and does not depend on the distance $\ox$ from the observer. In contrast, for the  $(\perp)$ scenario, the uncertainty in the measured separation radius $\oR^{\perp}$ becomes significant, and \textit{does} depend on the distance $\ox$.
The general orientation case will be addressed at the end of this section.

\subsection{Covariance Relations \& Notations}\label{Ssection: Some Definitions and notations}
To evaluate the physical uncertainties $\sigma_{\oR},\sigma_{\Opromeg}$, it is essential to first establish their relation to the fundamental uncertainties $\sigma_{\ox},\sigma_{\oy},\sigma_{\oz}$. This requires computing the \textit{covariance matrix} of $\bR$, denoted as $\K_{\bR\bR}$, for either of the configuration cases. For a three-dimensional vector $X$, the covariance matrix $\K_{XX}$—sometimes denoted as $\textbf{Var}(X)$—is defined component-wise as
\bea 
\left(\K_{X_iX_j}\right)_{ij}=\text{Cov}\left(X_i,X_j\right),\quad i,j =1,2,3.\nonumber
\eea
the diagonal terms are just $\sigma^2_{X_i}$, and the off-diagonal ones are the correlations between the vector components. 
Additionally, in our analysis, we will employ the generalized form of the covariance matrix, known as the \textit{cross-covariance matrix}, which is defined for two three-dimensional vectors $(X,Y)$ by its components:
\bea 
\left(\Sigma\left[X_i,Y_j\right]\right)_{ij}=\text{Cov}\left(X_i,Y_j\right), \quad i,j=1,2,3. \nonumber
\eea
This matrix satisfies several useful properties outlined in Appendix \ref{Appendix useful relations}. Also note that the covariance matrix $\K_{XX}$ is a special case of $\Sigma\left[XY\right]$ when $X=Y$.  
Let us now summarize shortly the notation we use here with $\hat{g},\hat{h}$ and $\hat{X},\hat{Y}$ random scalars/vectors operators respectively: 
\begin{itemize}
    \item $\sigma^2_{\hat{g}}$ or $\text{Var}(\hat{g})$ represent scalar variances.
    \item $\text{Cov}(\hat{g},\hat{h})$ represents scalar covariance.
    \item $\K_{\hat{X}\hat{X}}$ or $\textbf{Var}(\hat{X})$ denote covariance matrices.
    \item $\Sigma\left[\hat{X}\hat{Y}\right]$ represents the cross-covariance matrix.
    \item $(\ox,\oy,\oz)$ are space \textit{operators}, which should not be confused with direction unit vectors.
\end{itemize}

Let us now establish some useful relations (see Appendix \ref{appendix relations}) that relate the typical uncertainties of $\Opromeg$, $\oR$, and $(\ox,\oy,\oz)$ through the relations derived from \eqref{local frequency} (with all variables replaced by operators) and \eqref{Uncertainty Relation}, taking into account that $\hat{t} \sim \Opromeg^{-1}$.

\bea
|\sigma_{\Opromeg}|&=&\frac{3}{2}\sqrt{\frac{G_NM_T}{\AvR^5}}\sigma_{\oR},\label{std: omega to radius}
\\
\sigma_{\hat{t}}&\sim&\hat{t}^2\sigma_{\Opromeg}=\Opromeg^{-2}\sigma_{\Opromeg}=\frac{3}{2}\sqrt{\frac{\AvR}{G_NM_T}}\sigma_{\oR}, \label{std: time to radius}
\\
\sigma_{\ox}&\sim&\GothA\frac{\bar{x}}{\sigma_{\oR}},\quad
\sigma_{\oy}\sim\GothA\frac{\bar{y}}{\sigma_{\oR}},\quad
\sigma_{\oz}\sim\GothA\frac{\bar{z}}{\sigma_{\oR}},
\label{std: position to radius} 
\eea
where $\GothA:=\frac{l_p}{c}\sqrt{\frac{G_NM_T}{\AvR}}$.
It can be seen from here that only uncertainties coming along the $\ox$ direction will be important since its magnitude (the distance of a BBH) is much higher than the typical magnitude of $(\oy,\oz)$ (which scales as the separation of the BBH). 

When we determine the covariance matrix $\K_{\bR\bR}$ we will be able to evaluate $\sigma_{\oR}$ using the first-order Delta-Method from statistics (see, e.g., pp. 240-245 in \cite{statistical_Delta}). This method states that for a smooth function $f(\bR)$ (non-linear in $\bR$), its variance can be approximated as
\bea 
\text{Var}\left(f(\bR)\right)&\approx& \nabla f(\bR)^{\mathbf{T}}\K_{\bR\bR}\nabla f(\bR),
\neweqline
\nabla&:=&\left(\partial_{\bR_x},\partial_{\bR_y},\partial_{\bR_z}\right).
\eea
We thus apply this formula and derive for $\sigma_{\oR}$ in terms of the components of the matrix $\K_{\bR\bR}$: 
\bea\label{scalar radius from matrix}
\text{Var}(\oR)&=&\text{Var}\left(\sqrt{\bR_x^2+\bR_y^2+\bR_z^2}\right) \nonumber
\\
&\approx&\frac{1}{\oR^2}\left[\bR_x^2\text{Var}(\bR_x)+\bR_y^2\text{Var}(\bR_y)+\bR_z^2\text{Var}(\bR_z)\right]\nonumber
\\
&&+ \frac{2}{\oR^2}\left[\bR_x\bR_y\text{Cov}(\bR_x,\bR_y)+\bR_x\bR_z\text{Cov}(\bR_x,\bR_z)\right.\nonumber
\\
&&~~~~~~~~~ + \left.\bR_y\bR_z\text{Cov}(\bR_y,\bR_z) \right]
\eea
As we will explicitly see later, the off-diagonal covariance terms in $\K_{\bR\bR}$ (the 3rd and 4th lines in \eqref{scalar radius from matrix}) either vanish or remain negligibly small. Consequently, we arrive at a simplified relation between $\sigma^2_{\oR}$ and the diagonal components $(\K_{\bR\bR})_{gg}$ with $g\in\{x,y,z\}$. 

\subsection{Evaluation of Uncertainties}\label{ssection: evaluation of uncertainties}

We are now ready to evaluate the physical uncertainties $\sigma_{\oR},\sigma_{\Opromeg}$ through evaluating the covariance matrix $\textbf{Var}\left(\bR\right)$ for each of the configurations described earlier.

\subsubsection{Uncertainties for Face-on Case $(\parallel)$}

Based on our setup, by separating into the 3-dimension Euclidean coordinates \eqref{Euclidean coordinates}, the locations of each mass are given by 
\bea\label{coordinate of the parallel case}
\Oprr^{\parallel}_1 \cdot\unitx &=& \Oprr_2^{\parallel} \cdot\unitx:= \ox^{\parallel}, 
\neweqline
\Oprr^{\parallel}_2 \cdot\unity &=& -\Oprr^{\parallel}_1 \cdot\unity := \oy^{\parallel}=\oR^{\parallel}/2,
\eea
and equivalently for the $\oz$ axis, as these two axes are interchangeable, therefore, the uncertainties along these two axes are identical. Moreover, as we will see later for the $\hat{y}$ and $\oz$ axes, their uncertainty contributions are marginal, and the only significant component is along the $\ox$-axis. Because of this property our implicit assumption in \eqref{coordinate of the parallel case}, and for the rest of this paper, that the masses of the two bodies are equal (thus $\hat{y}_1=-\hat{y_2}$) would not change the final result. The inclusion of some deviations from  $\hat{y}_1=-\hat{y_2}$ will only result in differences in the uncertainty along $\oy,\oz$ which are marginal in any case. 

To proceed, we express the covariance matrix of the separation radius $\bR$:

\bea\label{std zero start} 
\textbf{Var}\left({\bR^{\parallel}}\right)&=&\textbf{Var}\left({\Oprr_2^{\parallel}-\Oprr_1^{\parallel}}\right)
\\
&=&\textbf{Var}(\Oprr^{\parallel}_1)+\textbf{Var}(\Oprr^{\parallel}_2)
\neweqline
&&-\Sigma\left[\Oprr^{\parallel}_1,\Oprr^{\parallel}_2\right]-\Sigma\left[\Oprr^{\parallel}_2,\Oprr^{\parallel}_1\right]\nonumber
\eea
Evaluating each variance, we first derive for $\text{Var}(\Oprr^{\parallel}_1)$:
\bea\label{Variance parallel stg1}
\textbf{Var}(\Oprr^{\parallel}_1)&=&\textbf{Var}(\ox\cdot \unitx-\oy\cdot \unity)
\neweqline
&=& 
\begin{pmatrix}
\text{Var}(\ox) & -\text{Cov}(\ox,\oy)\\
-\text{Cov}(\ox,\oy) & \text{Var}(\oy)
\end{pmatrix}.
\eea
Similarly, for $\textbf{Var}(\Oprr^{\parallel}_2)$ we obtain
\bea\label{Variance parallel stg2}
\textbf{Var}(\Oprr^{\parallel}_2)&=&\textbf{Var}(\ox\cdot\unitx+\oy\cdot\unity)
\neweqline
&=& 
\begin{pmatrix}
    \text{Var}(\ox)&\text{Cov}(\ox,\oy)\\
    \text{Cov}(\ox,\oy)& \text{Var}(\oy)
\end{pmatrix}.
\eea
Finally, evaluating the covariances in \eqref{std zero start}, we obtain for the first
\bea\label{Covariance parallel} 
\Sigma\left[\Oprr^{\parallel}_1,\Oprr^{\parallel}_2\right]&=& \Sigma\left[\ox\cdot\unitx-\oy\cdot\unity,\ox\cdot\unitx+\oy\cdot\unity\right]
\neweqline
&=&
\begin{pmatrix}
    \text{Var}(\ox) &  \text{Cov}(\ox,\oy)\\ 
    -\text{Cov}(\ox,\oy) &-\text{Var}(\oy)
\end{pmatrix},
\eea
and for the second: 
\bea\label{Covariance parallel stg2} 
\Sigma\left[\Oprr^{\parallel}_2,\Oprr^{\parallel}_1\right]&=& \Sigma\left[\ox\cdot\unitx+\oy\cdot\unity,\ox\cdot\unitx-\oy\cdot\unity\right]
\neweqline
&=&
\begin{pmatrix}
    \text{Var}(\ox) &  -\text{Cov}(\ox,\oy)\\ 
    \text{Cov}(\ox,\oy) &-\text{Var}(\oy)
\end{pmatrix}
\eea

Substituting \eqref{Variance parallel stg1}, \eqref{Variance parallel stg2}, \eqref{Covariance parallel}, and \eqref{Covariance parallel stg2} into \eqref{std zero start} we collect 
\bea
\textbf{Var}\left(\bR^{\parallel}\right)&=&2
\begin{pmatrix}
\text{Var}(\ox) & 0\\
0 & \text{Var}(\oy)
\end{pmatrix}
-2\begin{pmatrix}
    \text{Var}(\ox) & 0\\
0 & -\text{Var}(\oy)
\end{pmatrix}
\neweqline
&=&\begin{pmatrix}
    0 & 0\\
0 & 4\text{Var}(\oy)
\end{pmatrix}
\eea

Thus, using the last equality and the relation to $\sigma_{\oR}^2$ from \eqref{scalar radius from matrix}, we derive: 
\bea 
\sigma_{\oR^{\parallel}}^2\approx\frac{(\bR_{y}^{\parallel})^2}{\oR^2}\text{Var}(\bR_{y})=4\text{Var}(\oy).
\eea

With \eqref{std: position to radius} we can simplify $\text{Var}(\oy)$ and finally obtain for $\sigma_{\oR^{\parallel}}$:
\bea\label{Radius uncertainty explicit}
\sigma_{\oR^{\parallel}}^2&\approx&4(\GothA^{\parallel})^2\frac{1}{\sigma_{\oR^{\parallel}}^2}\bar{y}^2\Rightarrow
\neweqline
\sigma_{\oR^{\parallel}}^4&\approx&4(\GothA^{\parallel})^2\left(\frac{\AvR^{\parallel}}{2}\right)^2\Rightarrow
\neweqline
\sigma_{\oR^{\parallel}}&\approx&\sqrt{\frac{l_p}{c}}\left(G_NM_T\AvR^{\parallel}\right)^{\frac{1}{4}},
\eea
where $\GothA^{\parallel}:=\frac{l_p}{c}\sqrt{\frac{G_NM_T}{\oR^{\parallel}}}$. Therefore, the magnitude of the \textit{Relative Typical Uncertainty} (RTU) (which is a scalar) of the radius in Case (${\parallel}$) is given by (using \eqref{local frequency}):
\bea\label{std: Case (parallel) final}
\left|\frac{\sigma_{\oR^{\parallel}}}{\AvR^{\parallel}}\right|&\approx& \sqrt{\frac{l_p}{c}}\times\left(G_NM_T\right)^{1/4}\times\left(\AvR^{\parallel}\right)^{-3/4}
\neweqline
&=& \sqrt{\frac{l_p}{c}}\left(G_NM_T\right)^{1/4}\times\left[\left(G_NM_T\right)^{1/3}\times\bar{\omega}^{-2/3}\right]^{-3/4}
\neweqline
&=&
\sqrt{\frac{l_p}{c}\AvOme^{\parallel}}, 
\eea
which provides a dimensionless measure of the uncertainty along the $y$ direction parallel to $\bR^{\parallel}$ as measured by the observer. An important aspect of this result is that the RTU is independent of the distance $\ox$ from the observer. We discuss the physical reasoning of such behavior in Sec.~\ref{Observational sign}.
\\

\subsubsection{Uncertainties for Edge-on Case $(\perp)$} 
    \textbf{\textit{At maximal separation:}}

Before evaluating the uncertainties in the scenario illustrated in Fig.\ref{fig: edge on}, we look at the moment when $\phi=0$ for both simplicity and to enable physical understanding.  The event coordinates (again, along an Euclidean $(\ox,\oy)$ plane) for each mass are now those with maximal separation along the rotation plane. Without loss of generality, we assume that the mass with coordinates $\ox^{\perp}_2$ is the more distant object:
\bea\label{coordinates of the perpendicular case}
&&\Oprr^{\perp}_1\unity = \Oprr^{\perp}_2\unity = 0, 
\\
&&\Oprr^{\perp}_1\unitx = (\Oprr^{\perp}_2 - \bR^{\perp})\unitx := \ox_{1}^{\perp},
\Oprr^{\perp}_2\unitx := \ox^{\perp}_2,\nonumber
\eea

We begin by expressing the covariance matrix of the separation radius $\bR$: 
\bea\label{std one start}
\textbf{Var}\left(\bR^{\perp}\right)&=&\textbf{Var}\left(\Oprr^{\perp}_2-\Oprr^{\perp}_1\right)
\neweqline
&=& \textbf{Var}\left(\ox^{\perp}_1\cdot\unitx\right)+\textbf{Var}\left(\ox^{\perp}_2\cdot\unitx\right)
\neweqline
&&-\Sigma\left[\ox^{\perp}_1\cdot\unitx,\ox^{\perp}_2\cdot\unitx\right]-\Sigma\left[\ox^{\perp}_2\cdot\unitx,\ox^{\perp}_1\cdot\unitx\right]
\neweqline
&=& 
\begin{pmatrix}
    V_1&0\\
    0&0
\end{pmatrix} 
\eea
where we denote 
\beq
V_1=\text{Var}\left(\ox^{\perp}_1\right)+\text{Var}\left(\ox^{\perp}_2\right)-2\text{Cov}\left(\ox_2^{\perp},\ox^{\perp}_1\right). \nonumber
\eeq
Using the relation \eqref{scalar radius from matrix} we can deduce that the variance 
$\sigma_{\oR}^2$ is just the $xx$-component in the final 
line in \eqref{std one start}. 
Using now the known expressions for $\text{Var}(\ox^{\perp}_i)$ from \eqref{std: position to radius} and the property 
\bea
\text{Cov}(\ox^{\perp}_1, \ox^{\perp}_2) = \text{Var}(\ox^{\perp}_2) - \text{Cov}(\bR^{\perp}\cdot\unitx, \ox^{\perp}_2),\nonumber
\eea
we substitute into \eqref{std one start} and get 
\bea 
\sigma_{\oR^{\perp}}^2
\approx
\frac{(\GothA^{\perp})^2}{\sigma_{\oR^{\perp}}^2}\left(\left(\AvR^{\perp}\right)^2
-2\bar{x}^{\perp}_2\AvR^{\perp}\right)
+2\text{Cov}(\oR^{\perp},\ox^{\perp}_2),~~~
\eea
where $\GothA^{\perp}:=\frac{l_p}{c}\sqrt{\frac{G_NM_T}{\oR^{\perp}}}$. We also used that in this configuration $\bR^{\perp}\cdot\unitx=\oR^{\perp}$, and that
\beq
(\bar{x}_1^{\perp})^2+(\bar{x}_2^{\perp})^2-2(\bar{x}_{2}^{\perp})^2=\left(\left(\AvR^{\perp}\right)^2-2\bar{x}^{\perp}_2\AvR^{\perp}\right)\nonumber
\eeq

To keep the discussion compact, we now present our result for the RTU $|\sigma_{\oR^{\perp}}/\AvR^{\perp}|_{0}$ in Case $(\perp)$ with $\phi=0$, leaving the detailed calculations to Appendix \ref{appendix: Explicit Calculations of the uncertainties}. To derive this upper bound, we applied the following Cauchy-Schwarz inequality to the covariance term (see \cite{covariance_inequlities}):
\beq\label{covariance inequality}
\text{Cov}^2(A,B)\leq \text{Var}(A)\text{Var}(B). 
\eeq

Finally, we obtained the following result to leading order in $\AvR^{\perp}/\bar{x}^{\perp}_{2}$ (see \eqref{APP: STD final result}):
\bea\label{std: case(1) final} 
\left|\frac{\sigma_{\oR^{\perp}}}{\AvR^{\perp}}\right|_0\approx\sqrt{2\frac{v^{\perp}}{c}} \sqrt{ \frac{l_p\bar{x}^{\perp}_{2}}{(\AvR^{\perp})^2}}
=\frac{2}{3} \left|\frac{\sigma_{\Opromeg^{\perp}}}{\AvOme^{\perp}}\right|_{\phi=0}
,
\eea
where we have used the observed 
tangential velocity of the binary $v^{\perp} = \AvR^{\perp} 
\AvOme^{\perp}$. Unlike the Face-On Case 
\eqref{std: Case (parallel) final}, this expression explicitly depends on the 
distance of the system from the observer. This is a crucial property regarding observation, to be discussed in Sec.~\ref{Observational sign}.
\\

    \textbf{\textit{At arbitrary rotation angle:}}

We shall now use the results obtained in \eqref{std: case(1) final} to evaluate the relative uncertainties of $(\oR^{G},\Opromeg^{G})$ for a general angle $\phi$ of rotation as in Fig.\ref{fig: edge on}. We thus establish the following (as constructed by the observer) coordinate system for each mass:
\bea\label{coordinates of the general case}
&&\Oprr^{G}_1\unity = -\frac{\oy^G}{2}, \quad \Oprr^{G}_2\unity = +\frac{\oy^G}{2}, 
\neweqline
&&\Oprr^{G}_1\unitx= (\ox^{G}-\oR^{\perp}\cos{\phi}), \quad \Oprr^{G}_2\unitx=\ox^{G},
\eea
where $\pm\oy^G/2$ correspond to the locations of the two masses along the $\oy$-axis, and $\ox^G$ represents the location of the more distant object along the $\ox$-axis. 

Following the procedure outlined in \eqref{std one start}, we evaluate the covariance matrix and expand the cross-covariances:
\begin{widetext}
\bea\label{uncertainty general stg1}
\textbf{Var}\left(\bR^{G}\right)&=&\textbf{Var}\left(\Oprr^{G}_2-\Oprr^{G}_1\right)
\neweqline
&=&
\textbf{Var}\left((\ox^G-\oR^{\perp}\cos{\phi})\cdot\unitx-\frac{\oy^G}{2}\cdot\unity\right)
+\textbf{Var}\left(\ox^G\cdot\unitx+\frac{\oy^G}{2}\cdot\unity\right)
\\
&&-\Sigma\left[\ox^G\cdot\unitx+\frac{\oy^G}{2}\cdot\unity,(\ox^G-\oR^{\perp}\cos{\phi})\cdot\unitx-\frac{\oy^G}{2}\cdot\unity\right]\nonumber
-\Sigma\left[(\ox^G-\oR^{\perp}\cos{\phi})\cdot\unitx-\frac{\oy^G}{2}\cdot\unity,\ox^G\cdot\unitx+\frac{\oy^G}{2}\cdot\unity\right].\nonumber
\eea

Let us now evaluate each term separately:
\bea\label{general first var}
 \textbf{Var}\left((\ox^G-\oR^{\perp})\cdot\unitx-\frac{\oy^G}{2}\cdot\unity\right)
&=& 
 \begin{pmatrix}
     \text{Var}\left(\ox^G-\oR^{\perp}\cos{\phi}\right) &-\frac{1}{2}\text{Cov}\left(\ox^G-\oR^{\perp}\cos{\phi},\oy^G\right)
     \\
     -\frac{1}{2}\text{Cov}\left(\ox^G-\oR^{\perp}\cos{\phi},\oy^G\right) & \frac{1}{4}\text{Var}(\oy^G)
 \end{pmatrix}\nonumber
 \\
 &=&
 \begin{pmatrix}
   \text{Var}(\ox^G)+\cos^2\phi \text{Var}(\oR^{\perp})-2\cosp \text{Cov}(\ox^{G},\oR^{\perp})   & -V_2 \\
   -V_2 & \frac{1}{4}\text{Var}(\oy^G)
 \end{pmatrix},
\eea
where we denote
\beq
V_2=\frac{1}{2}\left(\text{Cov}(\ox^G,\oy^G)-\cosp \text{Cov}(\oR^{\perp},\oy^{G})\right). \nonumber
\eeq
For the second variance term, we get:
\bea\label{general second var}
\textbf{Var}\left(\ox^G\cdot\unitx+\frac{\oy^G}{2}\cdot\unity\right)
=
\begin{pmatrix}
  \text{Var}(\ox^G) & \frac{1}{2}\text{Cov}(\ox^G,\oy^G)
  \\
  \frac{1}{2}\text{Cov}(\ox^G,\oy^G) & \frac{1}{4}\text{Var}(\oy^G)
\end{pmatrix} ~~~~~
\eea
Next, we evaluate the cross-covariances, for the first we get:
\bea\label{general covariance}
\Sigma\left[(\ox^G-\oR^{\perp}\cos{\phi})\cdot\unitx-\frac{\oy^G}{2}\cdot\unity,\ox^G\cdot\unitx+\frac{\oy^G}{2}\cdot\unity\right]
&=&
\begin{pmatrix}
    \text{Cov}\left((\ox^G-\oR^{\perp}\cosp),\ox^G\right) & \frac{1}{2}\text{Cov}\left(\ox^G-\oR^{\perp}\cosp,\oy^G\right)\\
    -\frac{1}{2}\text{Cov}(\oy^G,\ox^G)&-\frac{1}{4}\text{Var}(\oy^G)
\end{pmatrix}\nonumber
\\
&=&
\begin{pmatrix}
    \text{Var}(\ox^G)-\cosp \text{Cov}(\oR^{\perp},\ox^G) & V_3\\
    -\frac{1}{2}\text{Cov}(\oy^G,\ox^G)&-\frac{1}{4}\text{Var}(\oy^G)
\end{pmatrix}
\eea
where we denote 
\beq
V_3=\frac{1}{2}\text{Cov}(\ox^G,\oy^G)-\frac{1}{2}\cosp \text{Cov}(\oR^{\perp},\oy^G). \nonumber
\eeq
And for the second cross-covariance, we note that they are equivalent up to exchange between the ${xy}$ and the ${yx}$ slots, hence the calculation is identical and we get: 
\bea\label{general covariance 2}
\Sigma\left[\ox^G\cdot\unitx+\frac{\oy^G}{2}\cdot\unity,(\ox^G-\oR^{\perp}\cos{\phi})\cdot\unitx-\frac{\oy^G}{2}\cdot\unity\right]
=
\begin{pmatrix}
    \text{Var}(\ox^G)-\cosp \text{Cov}(\oR^{\perp},\ox^G) & -\frac{1}{2}\text{Cov}(\oy^G,\ox^G)\\
    V_3
     &-\frac{1}{4}\text{Var}(\oy^G)
\end{pmatrix}~~~~~
\eea

\end{widetext}

Collecting \eqref{general first var}, \eqref{general second var}, \eqref{general covariance}, and \eqref{general covariance 2} into \eqref{uncertainty general stg1}, we find:
\bea\label{uncertainty general stg2}
\textbf{Var}(\bR)=
        \begin{pmatrix}
        \cospp \text{Var}(\oR^{\perp}) & \cosp \text{Cov}(\oR^{\perp},\oy^G)\\
            \cosp \text{Cov}(\oR^{\perp},\oy^G)& \text{Var}(\oy^G)
        \end{pmatrix}~~~~~
\eea

We now arrive at the final expression for the vector radius uncertainty for a given orientation relative to the observer. 
Notably, the special limiting case ($\perp$) is recovered when $\phi=0,\pi$.
To proceed from here and to get for the physical uncertainties $\sigma_{\oR},\sigma_{\Opromeg}$ we need to make use of \eqref{scalar radius from matrix}. At first, eq.\eqref{uncertainty general stg2} suggests we will need to evaluate the correlations as well as contributions along the $\hat{y}$ (and equivalently for $\oz$) axis. 
However, note that, as we got in \eqref{std: Case (parallel) final} and will see explicitly in Sec.\ref{Observational sign}, the term $\text{Var}(\oy)$ is marginal and is more than 10 orders of magnitude smaller than $\text{Var}(\oR^{\perp})$. 
Therefore, we can immediately neglect the ${yy}$ entry in \eqref{uncertainty general stg2}, and, regarding the $xy,yx$ entries, we can use again \eqref{covariance inequality} to say (taking the upper limit) that
\bea
\label{correlations marginal}
\text{Cov}(\oR^{\perp},\oy^{G})&\lesssim&\left(\text{Var}(\oR^{\perp})\text{Var}(\oy^G)\right)^{1/2}
\neweqline
&\ll&\text{Var}(\oR^{\perp}).
\eea
Therefore, we can relate $\sigma_{\oR^G}$ to $\textbf{Var}(\bR)$ in \eqref{uncertainty general stg2} simply by 
\bea\label{final relation scalar radius}
\sigma_{\oR^G}^2&\approx&\frac{\bR_{x}^2}{\oR}\left(\textbf{Var}(\bR)\right)_{xx}
\\
&=&\cos^2{\phi}\left(\cos^2{\phi}\text{Var}(\oR^{\perp})\right)=\cos^4{\phi}\text{Var}(\oR^{\perp}).\nonumber
\eea

To compute the total RTU over one binary cycle, we average by integrating \eqref{uncertainty general stg2} over $\phi$ in the domain $\phi\in [0,2\pi]$, thus incurring a factor of $3\pi/4$ compared to the result in  \eqref{std: case(1) final}.  This angular averaging reflects a statistical accumulation of variance (not a systematic drift) within a single cycle. Therefore, the total (averaged) accumulated uncertainty over one binary cycle is:
\beq\label{cycle uncertainty}
\left<\left|\frac{\sigma_{\oR^G}}{\AvR}\right|\right>_{\text{1cycle}} =\frac{3\pi}{4} \left|\frac{\sigma_{\oR^{\perp}}}{\AvR}\right|_0
\eeq

\subsubsection{Uncertainties for the case of general inclination} 

Evaluating the RTUs in both the $(\perp)$ and the $(\parallel)$ configurations, we observe that the dependence of the RTU on the distance from the observer is present only in case $(\perp)$, where gravitational wave detection is not possible (see \cite{Maggiore:2007ulw} around eq.(3.332)). Whereas for the $(\parallel)$ case, when the GW antenna response is expected to be maximal, the uncertainty remains constant and, as discussed in Sec.~\ref{Observational sign}, is marginal and not significant for observational purposes.

\begin{figure}[htbp]
    \centering
    \includegraphics[width=0.8\linewidth]{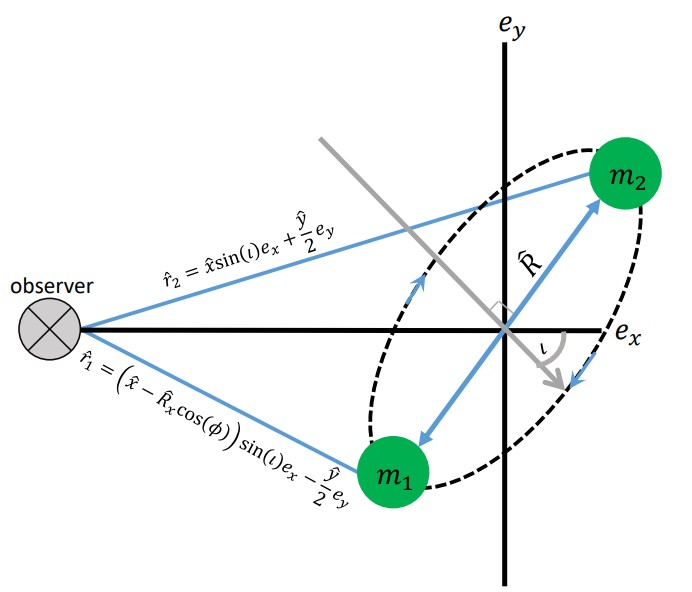} 
    \caption{Generic orientation}
    \label{fig: general ori}
\end{figure}

To have an observable that exhibits both a gain over distance dependence and detectability via gravitational waves, we consider a configuration where the BBH’s plane of rotation is inclined at an angle $\iota$ relative to the $\ox$ axis. This setup is illustrated in Fig.~\ref{fig: general ori}, where we define $\iota \in [0,\pi]$ as the angle between the normal to the plane of rotation and the $\ox$-axis.

The RTU of $(\oR^{G},\Opromeg^{G})$ that are statistically accumulated over one cycle in this configuration are analogous to those originated from the configuration in \eqref{coordinates of the general case}, except that each $\ox$ component of the coordinate vectors must be multiplied by $\sin{\iota}$. Any contributions coming from correlations of the uncertainty along $\ox$ with other axes will be marginal and can be dropped by the same arguments done previously in \eqref{uncertainty general stg2}, \eqref{correlations marginal}, and \eqref{final relation scalar radius}. Consequently, the final expression for the RTUs in a generic orientation follows from Eq.~\eqref{std: case(1) final}:
\bea\label{General Orientation uncertainties}
\left|\frac{\sigma_{\oR}}{\AvR}\right|_{\text{General}} =\left<\left|\sin{\iota}\frac{\sigma_{\oR^G}}{\AvR}\right|\right>_{\text{1cycle}} =\frac{3\pi}{4}\left|\sin{\iota}\frac{\sigma_{\oR^{\perp}}}{\AvR}\right|_0.~~~
\eea
As expected, when the rotation plane is edge-on ($\iota=\pi/2$) or face-on ($\pi=0$) we get the expected behavior of \eqref{cycle uncertainty} and \eqref{std zero start} respectively
\footnote{Unlike Eq.~\eqref{std zero start}, Eq.~\eqref{General Orientation uncertainties} appears to suggest that the relative uncertainty vanishes when $\iota=0$. However, the contribution along $\hat{y}$ persists in Eq.~\eqref{uncertainty general stg2}, implying that it remains present also in the general case.
In Eq.~\eqref{General Orientation uncertainties}, we have only highlighted the dominant contribution along the $\ox$-axis.}.

\section{Physical meaning \& Observation Significance}\label{Observational sign}
\subsection{Numerical Estimates}
Up to this point, we have analyzed how the RL in \eqref{Uncertainty Relation} affects the magnitude of uncertainty in measuring the frequency and separation radius of a BBH system during its inspiral stage. Our results indicate that this uncertainty depends on the orientation of the BBH’s plane of rotation relative to the observer. To better understand the difference in orientations, let us first insert some typical values for a BBH system into the RTU formulas.
As an example, using the separation radius at the end of the inspiral phase ($R_{\text{ISCO}}$) and the distance from the first detected event by the Advanced LIGO gravitational-wave detector  \cite{GW150914} (see \cite{adv-ligo} for the detector's description), we numerically evaluate the magnitude of the RTU for a face-on orientation, utilizing \eqref{std: Case (parallel) final}:  

\bea\label{case 0 numeric} 
\left|\frac{\sigma_{\oR^{\parallel}}}{\AvR^{\parallel}}\right|&\approx&\sqrt{\frac{l_p\AvOme^{\parallel}}{c}}=
\sqrt{\frac{v^{\parallel}l_p}{c\AvR^{\parallel}}}
\neweqline
&=&\sqrt{\frac{l_p}{\AvR^{\parallel}}}\approx\sqrt{\frac{10^{-35}m}{10^{5}m}}\sim 10^{-20}.
\eea

For comparison, using \eqref{std: case(1) final}, we derive the magnitude of the RTU in the edge-on orientation when $\iota=\pi/2$ and $\phi=0$:  
\bea\label{case 1 numeric}
\left|\frac{\sigma_{\oR^{\perp}}}{\AvR^{\perp}}\right|&\approx&\sqrt{2\frac{v^{\perp}}{c}} \sqrt{ \frac{l_p\bar{x}^{\perp}_{2}}{(\AvR^{\perp})^2}}\sim\sqrt{\frac{l_p\bar{x}^\perp_2}{(\AvR^{\perp})^2}}
\neweqline
&=&\sqrt{\frac{10^{-35}\times10^{26}m^2}{10^{10}m^2}}\sim 10^{-9}.\
\eea

It is apparent that when the orientation is face-on, the magnitude of the RTU is negligible compared to the edge-on one.
The reason behind this behavior lies in the nature of the RL 
effect described by \eqref{Uncertainty Relation}.
As highlighted in 
\cite{NonCommutative_Spacetime_Interpretation1}, and earlier in Sec.~\ref{ssection: Observation Preliminaries}, the RL effect is 
only observable when a coordinate transformation cannot alter the separation of 
observers.
In the face-on orientation, both black holes share the same distance along the line-of-sight ($\unitx$) axis, making them indistinguishable in the $\unitx$ direction concerning uncertainty: without momentum transfer along the $\unitx$ direction.
Consequently, the effect from the $\unitx$ separation is not measurable.
While there is a separation in the $\unity$ direction, proportional to the separation radius, the resulting RTU in \eqref{std: Case (parallel) final} remains marginal due to the small typical radius (on the order of tens of kilometers), hence the distance (from the observer) along $\unity$ is not enough to amplify the effect. 
In contrast, for an edge-on orientation, the RTU of the radius and angular frequency, given by \eqref{std: case(1) final}, strongly depends on the distance $\bar{x}^{\perp}_2$ between the detector and the BBH system. This dependence arises because, in this configuration, the two masses are located at different distances along the $\unitx$ direction and they transfer momentum along it, causing the uncertainty in measurement to be influenced by the $\unitx$ term in \eqref{Uncertainty Relation}.

Discussing in which orientations the RTU is significant or marginal, we now turn to identifying the properties of a BBH system that maximizes the RTU.  
We thus consider the general expression for the RTU over one cycle, given by \eqref{General Orientation uncertainties} (we included additional terms from \eqref{APP: STD final result}), highlighting different terms:  
\bea\label{Coloring general uncertainty}
\left|\frac{\sigma_{\oR}}{\AvR}\right|_{\text{G}} =
\frac{3\pi}{\sqrt{2}}\left|\sin{\iota}\right|\sqrt{\frac{v^{\perp}}{c}}\sqrt{\frac{l_p\bar{x}^{\perp}_{2}}{(\AvR^{\perp})^2}}
&&\left[1 - \frac{1}{\sqrt{2}}\frac{\AvR^{\perp}}{\bar{x}^{\perp}_{2}}\right.
\neweqline
&&~~~\left.+{\frac{1}{8}\left(\frac{\AvR^{\perp}}{\bar{x}^{\perp}_2}\right)^2}\right]
\nonumber\\
\!
\eea
The $v^{\perp}/c$ term in the equation reflects the \textit{relativistic} scale, indicating how close the system's tangential velocity approaches the speed of light. The $(\AvR^{\perp}/\bar{x}^{\perp}_{2})^{n=1,2}$ terms measure the system's size relative to the distance; these terms are typically marginal in any realistic astrophysical scenario. The $l_p\bar{x}^{\perp}_{2}/{(\AvR^{\perp})^2}$ term, however, is new and arises from the RL effect. This term is influenced by two main factors: the observer's distance from the system and the inverse-squared separation radius. As the observer’s distance increases, the effect of RL becomes more pronounced. Additionally, the system's energy increases as the separation radius decreases, amplifying the effect. 
Both the observer's distance and the separation radius are compared to the Planck length $l_p$. Consequently, the $l_p\bar{x}^{\perp}_{2}/{(\AvR^{\perp})^2}$ term measures how close we are to the Planckian region with a significant RL effect. We refer to this region as the Quantum Planckian Scale (QPS). This clarifies our choice of using a BBH system, as these systems exhibit the highest mass-to-radius ratio (compactness) and are thus ideal for exploring such effects.
The QPS represents a simultaneous limit of both the observer's distance and the system's energy. For a BBH system, the QPS can be expressed in terms of the observed radius $\bar{R}$ and the observed distance $\bar{x}$ as:
\bea\label{QPS Definition}
\frac{l_p\bar{x}_2}{\AvR^2}\sim 1 \Rightarrow \bar{x}_2 E_{\text{Orbit}}^2 \sim \frac{G_N^2m^4}{l_p}, 
\eea
where the orbital energy is given by $E_{\text{orbit}} = -\frac{G_N m^2}{2 \oR}$.
Equivalently, using units more customary in the gravitational-wave astrophysics community, we find that the RTU scales as
\bea
\left|\frac{\sigma_{\oR}}{\AvR}\right|_{\text{G}} \approx
    10^{-10}
    \left( \frac{M_\odot}{M}  \right)
    \sqrt{\frac{d_L}{1 \texttt{Mpc} }},
\label{RUT-gw}
\eea
where the deviation grows with the luminosity distance $d_L$ and shrinks with the black holes' typical mass $M$.

Finally, the $\left|\sin{\iota}\right|$ term represents the way the plane of rotation is oriented relative to the observer. At a first look, as discussed around \eqref{case 0 numeric} and \eqref{case 1 numeric}, to have the maximal RTU, we should take edge-on systems ($\iota=\pi/2$). 
However, since the way to observe the RTU in frequency is through gravitational waves (as we shall discuss shortly), and from the fact that gravitational waves are transverse waves, no detection is possible when the system is edge-on and the highest detection occurs for face-on orientation. Therefore, the preferable orientation angle $\iota$ is the one that maximizes both the gravitational wave detection and $\sin{\iota}$. As a simple example, we can consider the quadrupole radiation in leading order in $v/c$ treated in \cite{Maggiore:2007ulw} around eq.(3.332) (their $z$-axis is our $x$-axis), where the angular distribution is controlled by a $\cos{\iota}$, and therefore, the optimal orientation angle is $\iota=\pi/4$ so that $\sin{\iota}=\cos{\iota}=1/2$. 

\subsection{Observational Protocol \& accumulation over the inspiral}\label{ssection: protocol}
Assuming we had found a suitable BBH system according to the criteria stated earlier around eq.\eqref{Coloring general uncertainty}, an idealized observational procedure can be considered as follows:
\begin{itemize}
    \item[1.] Take an observation of the frequency at some determined early time $t_0$ of the inspiral such that $R(t=0)\ll R_\text{ISCO}$ and $|\sigma_{\omega}|\ll|\omega|$. This measurement will yield us what we defined as the \textit{local quantities} $(\omega_{GR}(t_0), R_{GR}(t_0))$, where the RTU is expected to be marginal.
    \item[2.] Construct from $(\omega_{GR}(t_0), R_{GR}(t_0))$ the expected form of $\omega_{GR}(t)$ at all times according to the classical theory of GR. 
    \item[3.] Sum or integrate over the RTU in \eqref{General Orientation uncertainties} along all the cycles inside the time frame $[t_0,t_f]$, where $t_{f}$ is the time where we can no longer consider the system to be in the inspiral stage $R_{GR}(t_f)=R_{\text{ISCO}}$. While summing, the radius and the frequency are those dictated by $\omega_{GR}(t), R_{GR}(t)$. Such a computation will result in the statistically accumulated RTU at all times $\sigma_{\omega}(t)$ representing the sum of variances contributed by each cycle.
    \item[4.] Take an observation of the system at times $t\in(t_0,t_f)$ and decide whether $\omega_{GR}(t)$ or $\omega_{GR}(t)+\sigma_{\omega}(t)$ better describes the measurements. 

\end{itemize}

Note that to evaluate the sum or the integration in stage \textbf{3}, we need to use an expression of $\omega(R)$ that increases with time to include radiation corrections from GR on the Kepler law \eqref{local frequency} (see e.g., eq.(4.25) in \cite{Maggiore:2007ulw}). 
Although the observation procedure we just described is in principle valid, and indeed, it could be used for a binary system located in a lab\footnote{A lab will have a small observational distance $x_2$, thus the RTU will be measurable only for energies near Planckian.} where such analysis can be done.

However, for a realistic application, and to have a high $x_2$ factor, any observation we perform must be on astrophysical systems through the analysis of emitted GW detected by one or more gravitational wave detectors.
The LIGO-Virgo-KAGRA Collaborations (LVK) \cite{adv-ligo,adv-virgo, KAGRA, Ushiba:2024lvn}, for example, already accompany each catalog of detections (as well as exceptional events) with a paper (or papers) detailing tests performed on the event(s) for deviations from GR \cite{LIGOScientific:2016lio, TheLIGOScientific:2016pea, LIGOScientific:2019fpa, gwtc2, LIGOScientific:2021sio}, in particular, tests of deviations during the inspiral stage, according to different imaginable post-Newtonian (PN) deviations \cite{Cornish:2011ys, Vitale:2016, Yagi:2016jml, Yunes:2013dva, Yunes:2010qb, Mehta:2022pcn, PhysRevD.90.064009}.
The crucial aspect of such observations is that tests of deviations from the classical GR are done through the observed allowed deviations in each order of the PN-expansion (see e.g., Fig.7 in \cite{LIGOScientific:2021sio}) of GR.
This framework exapnds the equations of GR or possible alternatives in orders of $\mathcal{O}\left(v/c\right)^{2n}$, where n is called the PN-order and $n\in[0,1/2,1/3/2,\ldots]$ with each increasing term revealing more information that comes from GR correction to Newtonian gravity. Concerning deviations from GR, the information that the GW measurements by LVK supply us are expressed as $\delta\phi_n$, which constrains how much the observation allows for deviating from GR in each PN order. 
Therefore, if we are to use the LVK constraints to evaluate whether our predictions for the RTU on the frequency of the BBH, we need first to cast the general result in \eqref{General Orientation uncertainties} into powers of $\frac{v}{c}$:
\bea\label{PN order} 
\left|\frac{\sigma_{\Opromeg}}{\bar{\omega}}\right|_{\text{General}} &\sim& F(x_2,l_p,G,M_T)\times \mathcal{O}\left(\sqrt{\frac{v}{R^2}}\right)
\neweqline
&\sim&F\times\mathcal{O}\left(\sqrt{v\times v^4}\right) \sim F\times\mathcal{O}(v^{5/2}),
\eea
where we denoted with $F(x_2,l_p,G,M_T)$ as all the constants in \eqref{General Orientation uncertainties}, and used (as a leading order approximation) the Kepler relation $v=R^{-1/2}$. 
In principle, we should have checked whether the magnitude of $F$ is allowed as a deviation in the PN order of $v^{5/2}$. Interestingly, such a power corresponds to $n=5/4=1.25$ in the PN expansion; a power that is \textbf{absent from} the PN expansion of GR, as well as from the theoretical alternatives commomly examined\footnote{The uniqueness of $\kappa$-GR can be seen here as it introduces a microscopic length-scale $\lp$ coupled directly to the time scale $\omega^{-1}$, whereas in other theories velocities, separations and frequencies must couple to one-another.}.
Hence, such deviations of "quarter-PN-orders" have (to our knowledge) not yet been directly constrained either with GW observations from LVK or other observations.
Unfortunately, this means that for now, we can not directly use existing published bounds to constrain our predictions by observational data;
we defer developing gravitational waveforms with quarter-PN modifications, and using them to asses bounds on such terms, to future work.
We do note that the bounds on the nearest orders, 1PN and 1.5PN, are similar, and of the order of $10\%$ compared to the GR expected values\footnote{As an illustration, in the Test of GR (TGR) analysis by the LVK \cite{LIGOScientific:2021sio} (see Table~VI therein), the sensitivity to deviations at each PN order $n$ is denoted by $\delta\phi_{2n}$. Typical current constraints for the lowest PN orders are at the $1\!-\!10\%$ level, e.g., $\delta\phi_{2} \simeq 0.05^{+0.09}_{-0.09}$ and $\delta\phi_{3} \simeq -0.03^{+0.06}_{-0.06}$.
} \cite{LIGOScientific:2021sio} - and so, for BBH systems as detected thus far, we expect 1.25PN deviations of the order of $10^{-10}$ to still be out of experimental reach.

We note also that equations  \eqref{General Orientation uncertainties} and  \eqref{Coloring general uncertainty} indicate additional PN deviations of orders $0.25,-0.75$, i.e. that gravity in the $\kappa$-Minkowski spacetime contains more modes not present in GR.
In other words, it tells us that if one establishes a consistent GR on the $\kappa$-Minkowski spacetime (such as the one we proposed in \cite{Our_Paper}), they should expect a PN expansion that contains powers like $n=1.25,0.25,-0.75$ in addition to the classical ones.    
We do note that these are suppressed by further $R/d_L$ factors.

\section{Summary and Outlook}\label{ssection: summary and discussion}
The $\kappa$-Minkowski spacetime \eqref{Kappa minkowski} has a central role in theories of QG phenomenology, as in DSR theories, and in general, as a NCST originating from some yet-unknown fundamental QG theory. 
In recent decades there have been growing efforts to extract observable phenomena that result from \eqref{Kappa minkowski}, such as time delays in the propagation of high energetic particles \cite{RL_Test_1, In_Vacua_Dis, RL_Test_2}, and generation of fields, such as gravitation \cite{Our_Paper}, 
see more in Sec.\ref{ssection: Introduction}.

A general property, that is independent e.g., in the choice of momenta basis used to implement DSR deformation, of the NCST \eqref{Kappa minkowski}, is the resultant uncertainty\footnote{This relies on a specific operator-valued realization of $\kappa$-Minkowski, but it does not depends on the choice of deformed symmetry within the DSR framework that preserves the same spacetime algebra, see Sec.\ref{ssection: Introduction} for more details.} \eqref{Uncertainty Relation} that was interpreted and understood in the past decade as the RL effect \cite{NonCommutative_Spacetime_Interpretation1, Relative_Locality}. The statement is that observers, observing the same event but from different distances will associate different spacetime coordination of the event. The crucial property is that the uncertainty is observable only if the discrepancy between the two observers is high enough, therefore the separation between them should be high. The reason is that they would not know what to compare with their notes to state that they got an uncertainty rather than just the classical answer, and since the uncertainty scales with the distance, the separation between them must be high. 

In this work, we opt to suggest an observable for the RL effect, advancing its theoretical stage into phenomenology. The initial reasoning was firstly to recognize that to compensate for the smallness of the Planck scale, the two observers should be separated by astrophysical distances, which is quite challenging to find at the moment. The second was to note that GW detection of BBH had reached a sufficient level of constraining deviations from GR and that BBH is far away and energetic. 
With these reasons in mind, we proposed here to consider the measurement made by \textit{one observer} at different \textit{times} of a BBH system at its inspiral stage, assuming, either as a 1st-order approximation or as a way to isolate the observational effect, that the internal dynamics of this BBH system remain undeformed by $\kappa$-GR. The observable effect of RL in this case turned out to be an uncertainty in the observed frequency (and in the separation radius) compared to the one predicted by the classical theory of gravitation. This uncertainty, as expressed for one cycle in \eqref{General Orientation uncertainties}, can be (by the right orientation angle) dependent on the observer's distance from the system and be observable by emitted GW. And, it depends on the energy of the binary, as time evolves and the separation radius shrinks, the uncertainty gets higher, therefore, we have the desired \textit{note comparing} between early times to later times when the uncertainty is marginal/important respectively. 

A crucial aspect concerning the observability of the derived effect is that the formula in \eqref{General Orientation uncertainties} describes the additional uncertainty in each cycle, which statistically accumulates\footnote{As the sum of the corresponding variances. This is in contrast to a classical, deterministic deviation, which would result in a systematic drift.}. Therefore, to evaluate the expected uncertainty at the end of the inspiral, one should sum over all the uncertainties of the relevant cycles, amplifying the uncertainty at late inspiral times into something observable in principle. 
Because constraints on deviations from GR via observation of GW are often given in terms of allowed deviations in each order $n$ in $(v/c)^{2n}$ of the PN-expansion \cite{Blanchet:1994ez,Arun:2006yw, Arun:2006hn,Yunes:2009ke,Mishra:2010tp,Testing_GR_2,Will:2014kxa,LIGOScientific:2016lio,TheLIGOScientific:2016pea,LIGOScientific:2019fpa,gwtc2,LIGOScientific:2021sio}, we cast the form of the uncertainty \eqref{General Orientation uncertainties} into an order of $(v/c)$. 
Interestingly, we derived in \eqref{PN order} the (relative) typical uncertainty in the frequency scales as $(v/c)^{5/2}$, resulting in the corresponding $n=1.25$ in the PN-expansion, an order without available constraint data from the LVK collaboration since such a PN-order does not exist in GR (only $n=0,0.5,1,1.5,\ldots$, and $n=0.5$ which is easily treated in similar fashion). 
Unfortunately, this means that for now, we can not constrain our prediction via existing constraints from GW observation and that we should wait for future analysis of the GW data also for PN-orders like $n=1.25,0.25,-0.75,\ldots$ to have constraints on allowed deviations. 
Nevertheless, assuming our results are sound, we can learn from the PN order of the frequency uncertainties that a gravity theory on the \eqref{Kappa minkowski} spacetime introduces powers in the PN-expansion that are not present in GR, and therefore, to probe the validity of \eqref{Kappa minkowski} with gravity, one should study constraints that come from non-classical PN orders that until now were left unstudied.  

As future work on the subject, we consider checking whether a gravity theory on \eqref{Kappa minkowski} (like the one we developed in \cite{Our_Paper}) will indeed produce the non-trivial PN orders like $n=1.25$ as we found here, enabling also the assessment of what are the implications of including \textit{generation} induced modifications. And, to be able to compare \eqref{General Orientation uncertainties} with observations, we hope that we or others will turn an endower to evaluate the constraints on the $n=1.25$ PN order by constructing a suitable waveform and using the data from the LVK collaboration, and future gravitational wave detectors. This could be done either by applying the PN expansion to $\kp$-GR, where a 1.25PN is expected to arise and could be analytically incorporated into waveform models using standard techniques,
    similarly to the Effective-One-Body family of waveforms \cite{Bohe:2016gbl,Cotesta:2018fcv,Cotesta:2020qhw,Dietrich:2019kaq,Matas:2020wab} used for the flexible theory-independent (FTI) method \cite{Mehta:2022pcn,LIGOScientific:2018dkp} parameterizing deviations from GR by PN frequency orders, 
    or more phenomenologically, by incorporating a 1.25 PN
    correction into existing templates such as the IMRPhenom family \cite{Pratten:2020ceb, Garcia-Quiros:2020qpx, Colleoni:2023czp}, which should be in principle adaptable to such a modification,
    as done for the TIGER tests \cite{{Li:2011cg, Agathos:2013upa, Meidam:2017dgf, Colleoni:2023czp, roy:2024tiger}}.
    Both of these infrastructures and waveform families are in common use by the LVK, and already include modifications in half-powers of the PN parameter ($\omega^{1/3}$, equivalent to $v/c$), so quarter-powers require changing the code, but in straightforward manners.

We note in particular that the prospects for measurement increase with increasing distance and with decreasing BH mass - which prompts a challenge of detecting small black holes, at ultra-high high gravitational wave frequencies, out to very large cosmological distances - a challenge indeed.
Finally, we note that the greatest impediment to such measurements is the \"uber-astronomical smallness of the Planck length scale, i.e., the observation (rel. equations (\ref{QPS Definition},\ref{RUT-gw})) that the farthest cosmological distance observed, $O(10 \texttt{Gpc})$, is only $\sim10^{20}$ times larger than typical stellar black holes, while the Planck length is about $10^{40}$ times smaller than such black holes - and it is the square root of the quotient of these two ratios which gives the order of the expected deviation. 
Even with upcoming GW-detector technologies, such as the planned space interferometer LISA \cite{LISA:2017pwj}, the magnitude estimated in \eqref{case 1 numeric} will remain beyond direct sensitivity.
Similar considerations apply to ground-based upgrades of the current facilities and to the planned Einstein Telescope \cite{ET:2019dnz}, for which the required parameter regime (smaller black holes at extreme distances) is challenging to access astrophysically. 
Nevertheless, it is not excluded that future detector generations or improved analysis methods could bring such effects within reach, and detailed simulations following the launch of LISA and/or construction of ultra-high-frequency facilities will be needed to assess the prospects. 
Specifically, we note that, according to \eqref{RUT-gw}, for a BBH system that is on the one hand very low in mass and on the other hand very nearby, the effect could reach order unity and thus become observable (e.g., for $M \sim 10^{-10}M_{\odot}$ and $d_L \sim \mathrm{pc}$ the deviation is measurable).
Such scenarios are admittedly quite specific, and would most naturally be associated with primordial black holes (see, e.g., \cite{PrimeBH1,PrimeBH2}), and should be investigated in future work. More generally, even if such extreme configurations are not realized in nature, the situation could change if Relative Locality effects of the kind we considered here were to arise at larger length scales or lower energies. In that case, the same framework could provide testable predictions for future gravitational-wave observations.

\section*{Acknowledgments}
We acknowledge support from the US-Israel Binational Science Fund (BSF) grant No. 2020245 and the Israel Science Fund (ISF) grant No. 1698/22. We want to thank Yarden Shani,  Nadav Shnerb, David Kessler and Jonathan Ruhman for helpful discussions,
as well as the anonymous referee.

\appendix

\section{Useful properties of the Covariance}\label{Appendix useful relations}
For the cross-covariance matrix $\Sigma$, the following holds for any (random) $m$-dimensional vectors $X,Y,X_1,X_2$, constant (of dimensions $m\times m$) matrices $\mathbf{A,B}$, and constant $m$-dimensional vectors $\mathbf{a,b}$: 
\bea 
\Sigma_{XY}&=&\Sigma_{YX}^{\mathbf{T}},
\neweqline
\Sigma_{X_1+X_2,Y}&=&\Sigma_{X_1,Y}+\Sigma_{X_2,Y}, 
\neweqline
\Sigma_{\mathbf{A}X+\mathbf{a},\mathbf{B^{T}}Y+\mathbf{b}}&=&\mathbf{A}\Sigma_{XY}\mathbf{B}.
\eea

And, we also have the following properties of the covariance matrix $\textbf{Var}(X)=\K_{XX}=\Sigma_{XX}$:
\bea
\K_{XX}&=&\K_{XX}^{\mathbf{T}}, 
\neweqline
\textbf{Var}(X\mathbf{A}+\mathbf{a})&=&\mathbf{A}\textbf{Var}(X)\mathbf{A^{T}},
\neweqline
\textbf{Var}(X+Y)&=&\K_{XX}+\K_{YY}+\Sigma_{XY}+\Sigma_{YX}.
\eea

\section{Relations for Sec.~\ref{Ssection: Some Definitions and notations}}\label{appendix relations}
In the following, we derive the results stated in (\eqref{std: omega to radius}, \eqref{std: position to radius}, \eqref{std: time to radius}). For instance, consider the relation between 
$\sigma_{\hat{t}}$ and $\sigma_{\Opromeg}$, we begin by noting that 
$\hat{t}^{-1}\sim\Opromeg$, then, by Taylor-expanding 
$\Opromeg(\bar{t}+\sigma_{\hat{t}})$, we derive:
\bea\label{std relations 1}
\Opromeg(\bar{t}+\sigma_{\hat{t}})&=&\bar{\omega}(\bar{t})+\frac{d\Opromeg}{d\hat{t}}\bigg|_{\bar{t}}\sigma_{\hat{t}} \Rightarrow
\neweqline
\sigma_{\Opromeg}&=&\frac{d\Opromeg}{d\hat{t}}\bigg|_{\bar{t}}\sigma_{\hat{t}}=-\frac{\sigma_{\hat{t}}}{\bar{t}^2}
\eea 
From \eqref{local frequency} we know the relation between $\oR$ and $\Opromeg$; following a similar calculation done previously, we derive: 
\bea\label{std relations 2} 
\sigma_{\Opromeg}&=&\frac{d\Opromeg}{d\oR}\bigg|_{\AvR}\sigma_{\oR} =-\frac{3}{2}\sqrt{\frac{G_NM_T}{\AvR^5}}\sigma_{\oR},
\\\label{std relations 3}
\sigma_{\hat{t}}&=&\frac{\sigma_{\Opromeg
}}{\bar{\omega}^2}=\frac{3}{2}\sqrt{\frac{\AvR}{G_NM_T}}\sigma_{\oR}.
\eea

\section{Explicit Calculations of the uncertainties}\label{appendix: Explicit Calculations of the uncertainties}
We now outline the calculation that led to the result of the RTU in Case $(\perp)$ as appeared in \eqref{std: case(1) final}. In the body of the work, we derived the following 
\bea\label{APP STD I} 
\sigma_{\oR^{\perp}}^2
=
\frac{(\GothA^{\perp})^2}{\sigma_{\oR^{\perp}}^2}\left(\left(\AvR^{\perp}\right)^2
-2\bar{x}^{\perp}_2\AvR^{\perp}\right)
+2\text{Cov}(\oR^{\perp},\ox^{\perp}_2).~~~
\eea
By use of the inequality \eqref{covariance inequality} (we take the maximal value) we can evaluate $\text{Cov}(\oR^{\perp},\xtwo)$ in terms of known typical uncertainties, 
\bea\label{APP STD II}
\text{Cov}(\oR^{\perp},\xtwo)&\approx&\sigma_{\oR^{\perp}}\sigma_{\xtwo} =\sigma_{\oR^{\perp}}\GothA^{\perp}\frac{\bar{x}^\perp_2}{\sigma_{\oR^{\perp}}}
\neweqline
&=&\sqrt{\text{Var}(\oR^{\perp})\text{Var}(\xtwo)},
\eea
we used \eqref{std: position to radius}. Inserting \eqref{APP STD II} into \eqref{APP STD I}, we get 
\bea 
\sigma_{\oR^{\perp}}^2&\approx&\GothA^{\perp}\left[\frac{\GothA^{\perp}}{\sigma_{\AvR^{\perp}}}\left((\AvR^{\perp})^2
-2\bar{x}^\perp_2\AvR^{\perp}\right)+2\bar{x}^\perp_2\right].~~~
\eea
Which results in the following quadratic equation in $\sigma_{\oR^{\perp}}^2:=S$,
\bea
0=S^2-2\GothA^{\perp}\bar{x}^\perp_2 S -(\GothA^{\perp})^2\left((\AvR^{\perp})^2-2\bar{x}^\perp_2\AvR^{\perp}\right),~~~
\eea 

\begin{widetext}
The two solutions for such an equation are, 
\bea
S^2&=&\GothA^{\perp}(\bar{x}^\perp_2\pm\left[(\bar{x}^\perp_2)^2+\AvR^{\perp}(\AvR^{\perp}-2\bar{x}^\perp_2)\right]^{\frac{1}{2}})
=\GothA^{\perp}\bar{x}^\perp_2\left(1\pm\left[1+\frac{\AvR^{\perp}}{(\bar{x}^\perp_2)^2}(\AvR^{\perp}-2\bar{x}^\perp_2)\right]^{\frac{1}{2}}\right)
\approx
\GothA^{\perp}\bar{x}^\perp_2\left(1\pm\left(1+\frac{\AvR^{\perp}}{2(\bar{x}^\perp_2)^2}(\AvR^{\perp}-2\bar{x}^\perp_2)\right)\right),~~~~~~
\eea
we used the fact that the ${\AvR^{\perp}}/({\bar{x}^\perp_2)^2}$ term is smaller than unity in any astrophysical BBH, utilizing the Taylor expansion. Taking the $(+)$ in the above expression (for a real solution), we derive for the typical uncertainty $\sigma_{\AvR^{\perp}}$, 
\bea 
\sigma_{\AvR^{\perp}}
\approx
\sqrt{2\GothA\bar{x}^\perp_2}\left[1-\frac{\AvR^{\perp}}{4(\bar{x}^\perp_2)^2}(2\bar{x}^\perp_2-\AvR^{\perp})\right]^{\frac{1}{2}}
\approx
\sqrt{2\GothA\bar{x}^\perp_2}\left(1-\frac{\AvR^{\perp}}{8(\bar{x}^\perp_2)^2}(2\bar{x}^\perp_2-\AvR^{\perp})\right)
= \sqrt{2\GothA\bar{x}^\perp_2}\left[1-\frac{\AvR^{\perp}}{4\bar{x}^\perp_2}+\frac{1}{8}\left(\frac{\AvR^{\perp}}{\bar{x}^\perp_2}\right)^2 \right].
\eea
Using the definition of $\GothA$ and of the angular frequency \eqref{local frequency}, we can rearrange the last result to derive the RTU,
\bea\label{APP: STD final result}
\frac{\sigma_{\AvR^{\perp}}}{\AvR^{\perp}}&=&
\sqrt{\frac{2l_p\bar{x}^\perp_2}{c}}\Biggr[\left(\frac{\bar{\omega}}{\AvR^{\perp}}\right)^{\frac{1}{2}}
-\left(\frac{\bar{\omega}\AvR^{\perp}}{2(\bar{x}^\perp_2)^2}\right)^{\frac{1}{2}}+\left(\frac{\bar{\omega}(\AvR^{\perp})^3}{(\bar{x}^\perp_2)^4}\right)^{\frac{1}{2}} \Biggr] 
=\sqrt{2\frac{v^{\perp}}{c}}\sqrt{\frac{l_p\bar{x}^\perp_2}{(\AvR^{(\perp)})^2}}\left[1-\frac{1}{\sqrt{2}}\frac{\AvR^{(\perp)}}{\bar{x}^\perp_2}+\frac{1}{8}\left(\frac{\AvR^{(\perp)}}{\bar{x}^\perp_2}\right)^2\right].
\eea
In the last equality we rearranged a bit and used the tangential velocity $v^{\perp}:=\bar{\omega}\AvR^{\perp}$. 

\end{widetext}

\end{document}